# The Coulomb Hole of the Ne Atom

Mauricio Rodríguez-Mayorga,[a, b] Eloy Ramos-Cordoba,*[a] Xabier Lopez,[a] Miquel Solà,[b] Jesus M. Ugalde,[a] and Eduard Matito*[a, c]

We analyze the Coulomb hole of Ne from highly-accurate CISD wave functions obtained from optimized even-tempered basis sets. Using a two-fold extrapolation procedure we obtain highly accurate results that recover 97% of the correlation energy. We confirm the existence of a shoulder in the short-range region of the Coulomb hole of the Ne atom, which is due to an internal reorganization of the K-shell caused by electron correlation of the core electrons. The feature is very sensitive to the quality of the basis set in the core region and it is not exclusive to Ne, being also present in most of second-row atoms, thus confirming that it is due to K-shell correlation effects.

## 1. Introduction

Electron correlation remains as a central issue for the physico-chemical description of the electronic structure. Its study often provides physical insights to develop new computational methods to tackle the electronic structure of molecules. The primitive description provided by the Hartree-Fock (HF) wave function has been improved by consideration of different types of electron correlation, such as dynamic and nondynamic correlation, in the so-called post-HF methods as well as in methods that do not employ wave functions, such as the density and reduced-density matrix functional theories (DFT[35] and RDMFT[9,38,39,49]). The improvement of computational methods, the correct choice of a computational protocol to address a given problem, and our understanding of the electron correlation, hinge on the development of appropriate descriptors of electron correlation.[15,30,41–45,52,62] Lately, our efforts have concentrated in this direction, resulting in the development of simple electron correlation descriptors capable of separating dynamic and nondynamic correlation.[42,44,62]

The Coulomb hole stands among the classical descriptors that are used to study electron correlation due to its conceptual simplicity and its connection with the electron-electron interaction energy.[1,2,12,33] The Coulomb hole provides a practical picture of how the electron correlation affects the interelectronic separation. Namely, it reflects the change of the electron-electron distance distribution upon the inclusion of electron correlation. From this quantity the correlation effects on the average interelectronic distance, its variance, and the electron-electron repulsion are easily assessed. The topological features of the Coulomb hole have also been studied, leading to some relevant conclusions about the nature of electron correlation.[10,17,60] Some of us have also recently used the long-range part of the Coulomb hole to characterize van der Waals interactions.[61,62]

In this work, we analyze a key feature of the Coulomb hole of the Ne atom that, thus far, has been largely ignored by many quantum mechanics practitioners. In 1969, Bunge and co-workers identified a shoulder structure in the short-range part of the Coulomb hole of the Ne atom,[37] which was corroborated by Cioslowski and Liu thirty years later.[24] Bunge attributed this peculiarity to the K-shell electrons, whereas Cioslowski did not comment on this feature. We have found that the shoulder is very sensible to the quality of the basis sets employed in the calculation, turning into a mininum or vanishing depending on the basis set. In order to confirm the presence of the shoulder we have performed CISD and FCI calculations employing large optimized even-tempered basis sets, which provide energy estimates that compare well with the most accurate values obtained by Bunge.[4–6,37] Our results provide a thorough study on the origin of the shoulder, identifying the causes that are responsible for its presence. Finally, we prove that this feature is not exclusive to Ne atom.

## 2. Methodology

There are mainly three different ways to define correlation holes: McWeeny's,[33] Ros',[50] and Coulson's.[12] The former is statistically motivated and it does not employ reference wave functions, whereas the other two use HF as the uncorrelated reference. In this work, we are concerned with Coulson's definition, which is connected with an experimental observable, the X-ray scattering intensity. Coulson's Coulomb hole is obtained from the difference between the exact and the HF intracule densities. The radial intracule density provides a distribution of the electron-electron distances,

$$I(u) = \iint d\mathbf{r}_1 d\mathbf{r}_2 \, n_2(\mathbf{r}_1, \mathbf{r}_2) \delta(u - r_{12}), \tag{1}$$

where $n_2(\mathbf{r}_1, \mathbf{r}_2)$ is the pair density and $r_{12}$ is the module of the intracule coordinate, $\mathbf{r}_{12} = \mathbf{r}_1 - \mathbf{r}_2$.

The X-ray scattering intensity is essentially determined by the Fourier-Bessel transform of the intracule pair density[58,59] and it is employed in the study of elastic and inelastic scattering of electrons.[37] The total X-ray scattering intensity for short wavelengths is actually governed by the value of the intracule at the coalescence points, $I(0)$.[57]

The difference between the exact pair density and an uncorrelated reference, represents the change in the electron pair distribution upon the introduction of electron correlation. Coulson's Coulomb hole[12] sets HF intracule pair density as the uncorrelated reference,

$$h_C(u) = I(u) - I_{HF}(u) \qquad (2)$$

giving negative (positive) values for the interelectronic separations $u$ that are increased (decreased) upon the inclusion of correlation. The integration of $h_C(u)$ over $u$ gives zero.

Since the quality of the basis set is crucial for the description of the holes at short electron-electron distances, we have generated an optimized set of basis functions. The optimization of the basis sets employs an analogous procedure to the one developed elsewhere.[29] This procedure has been successfully used to generate highly-accurate basis functions to test model systems and calibrate a number of methods.[8,29,48,49,56] First of all, a family of uncontracted basis sets consisting of spherical Gaussian primitives is constructed by selecting the optimized exponents that minimize the CISD energies (the coefficients that multiply the primitives are equal to 1 and do not enter the optimization procedure). From these values, the complete-basis set (CBS) estimate of Ne CISD energies are obtained by a two-fold extrapolation procedure.

The family of basis sets employs functions with exponents $\zeta_{L,N}^k$ that are even-tempered[51] according to the expression

$$\zeta_{L,N}^k = \alpha_{L,N} [\beta_{L,N}]^{k-1}, \quad 1 \leq k \leq N. \qquad (3)$$

Each basis set is characterized by the maximum angular momentum, $L$, and the number of basis functions for each function type, $N$. For instance, 6SP ($L=1$, $N=6$) basis set consists of six groups of functions containing one S and three P functions ($p_x$, $p_y$ and $p_z$) sharing the same exponent. The exponent assigned to each group is given by $k$ in Eq. 3, which runs from 1 to $N$. $\alpha_{L,N}$ and $\beta_{L,N}$ are, therefore, unique for each basis set and determined by minimization of the CISD energy of Ne with a simplex method (minimal accuracy $10^{-7}$ a.u.). The family includes basis sets with angular momentum between 0 and $L$ ($1 \leq L \leq 4$) and involve equal numbers $N$ ($6 \leq N \leq 16$) of spherical Gaussian primitives with exponents $\zeta_{L,N}^k$, giving rise to 44 different basis sets.

The computed energies $E_{L,N}$ have been extrapolated to the respective $N \to \infty$ limits $E_L$ by fitting the actual energy values for $N=12$, 13, 14, 15 and 16 with the double-exponential expression

$$E_{L,N} = E_L + a_L e^{-\alpha_{L,N} N} + b_L e^{-\beta_{L,N} N}, \qquad (4)$$

which generalizes the Dunning extrapolation.[64] The resulting system of five non-linear equations has been solved analytically with Mathematica[63] employing the Ramanujan algorithm.[40]

In turn, the estimates $E_L$ have been extrapolated to the respective CBS limits $E$ by fitting the values of $E_L$ for $L=2$, 3, and 4 to the expression[19,24,25]

$$E_L = E + \frac{B}{[L+1]^3}. \qquad (5)$$

These extrapolations, $E_L$ and $E$, provide lower-energy estimates of the total energy that are not variational.

In the case of HF, the energy results are almost converged using only S and P basis functions. Therefore, we take the SP-energy limit as a good estimate of the CBS-extrapoled result. The numerical estimate is obtained from $N=16$, 17, 18, 19 and 20 calculations applying the fitting of Eq. 5.

The full-configuration interaction (FCI) calculations have been carried out with a modified version of the FCI program of Knowles and Handy[43] and the CISD calculations have been performed with Gaussian.[44] The calculations of the second-order reduced density matrices (2-RDM) have been calculated from the FCI/CISD expansions coefficients using the in-house DMN code.[45,46] The radial intracule density was computed with the in-house RHO2_OPS code,[47] which uses the algorithm proposed by Cioslowski and Liu.[48]

## 3. Results

### 3.1. Benchmark Data

Following the procedure described in the previous section we have obtained a CISD extrapolated energy of $-128.9254609$ a.u., which represents an energy lowering of $-0.0143843$ a.u. with respect to the best variational estimate, $E_{4,16}$ (see Table 1). These results compare well with the best non-relativistic FCI estimate available in the literature, $-128.937588$ a.u.[5]

Our CISD SPDF-energy limit, $-128.8984284$ a.u., is in good agreement with the FCI value $-128.897 \pm 0.002$ a.u. calculated by Bunge.[21] This and the other partial waves reported in Table 1 are also in accord with the second-order correlation energies of Lindgren and Salomonson.[28]

Our extrapolated HF energy, $-128.547100$ a.u., which also corresponds to the SP-energy limit, is in excellent agreement with the numerical HF results, $-128.547098$ a.u., reported elsewhere.[50] Our best CISD estimate of the correlation energy is, therefore, −0.378361 a.u., which represents 97% of the correlation energy of Ne.[1] Our best variational estimate of the correlation energy, based on the CISD/16SPDFG calculation (including 400 basis functions), recovers 93% of the correlation energy. Our calculations on the angular and the radial correlation[27] indicators of Kutzelnigg[26] show no qualitivative improvement in the description of correlation beyond the 11SPDF basis set (see Figures S1 and S2) and Bunge and coworkers report very small effects upon introduction of the triple and quadruple excitations (less than 0.01% change on the density).[3] Therefore, we conclude that our CISD calculations provide a satisfactory description of electron correlation in Ne.

Table 1. CISD energies (a.u.) for the basis set family developed in this work and the corresponding partial waves.

| N | $E_{1,N}$ | $E_{2,N}$ | $E_{3,N}$ | $E_{4,N}$ |
|---|---|---|---|---|
| 5 | −127.8146757 | −127.9311638 | −127.9638254 | −127.9755176 |
| 6 | −128.3326841 | −128.4543077 | −128.4887366 | −128.5010355 |
| 7 | −128.5692718 | −128.6923709 | −128.7272631 | −128.7395973 |
| 8 | −128.6612022 | −128.7855792 | −128.8209903 | −128.8335145 |
| 9 | −128.6987672 | −128.8240698 | −128.8598984 | −128.8725885 |
| 10 | −128.7176209 | −128.8433421 | −128.8794251 | −128.8922251 |
| 11 | −128.7265782 | −128.8526131 | −128.8888882 | −128.9017858 |
| 12 | −128.7304973 | −128.8567898 | −128.8932237 | −128.9062091 |
| 13 | −128.7323966 | −128.8588808 | −128.8954410 | −128.9085010 |
| 14 | −128.7334758 | −128.8600836 | −128.8967326 | −128.9098489 |
| 15 | −128.7340477 | −128.8607425 | −128.8974582 | −128.9106198 |
| 16 | −128.7343430 | −128.8611063 | −128.8978764 | −128.9110766 |
| ∞ | −128.7346499 | −128.8615534 | −128.8984284 | −128.9117007 |

We have also explored the convergence of certain properties related to the Coulomb hole with the size of the basis set. Our results indicate that the average interelectronic distance and its variance are much more affected by the number of basis functions than by the inclusion of functions of large angular momentum. In this respect, the use of 9SP basis functions provides a reasonable description of these indicators (see Figures S3 and S4). For this reason, we have chosen the CISD/9SP wave function to provide a qualitative explanation of the Coulomb hole in Ne atom. In a number of selected cases, analysis with larger basis sets have been performed to confirm our conclusions.

### 3.2. The Coulomb Hole of the Ne Atom

In his seminal paper, Bunge[23] reported a small shoulder of the Coulomb hole of Ne in the short interelectronic distances domain that he attributed to the electron correlation within the $K$-shell. This calculation was based on a FCI wave function that yield an electronic energy of −128.8602 a.u. and, thus, only retrieved 85% of the correlation energy.[6] Thirty years later, Cioslowski and Liu confirmed this result using 2-RDMs obtained from energy derivatives of MP2 calculations with a non-optimized even-tempered basis set of 50 functions ($20s10p$).[11] We have tried to reproduce the results of Bunge and Cioslowski and have encountered a major difficulty choosing the appropriate basis set. We have performed over hundred CISD calculations (and some FCI calculations as well) using different basis with and without the frozen core approximation, finding that the shoulder is only reproduced in about half of the cases (see Tables S1 and S2). No frozen-core calculation could reproduce the shoulder structure regardless the size of the basis set, supporting the idea that this feature, if not an artifact due to inaccurateness of the wave function, is a result of the correlation of the core electrons. The basis set families show similar results among its members. Pople's 6-311G and larger basis of this family as well as the first family of basis sets developed by Dunning ($nZ$) and the core correlated-consistent basis sets cc-pCV$n$Z display the shoulder structure.[14,65] Conversely, the family of correlated-consistent basis sets of Dunning[13] (cc-pV$n$Z) and the series of basis sets of Petersson[46] ($n$ZaP) cannot reproduce the shoulder structure (see Figure 1 for some examples).

In order to solve this controversy, we have built a series of even-tempered basis sets following the procedure described above. For all these basis sets, regardless the size, the shoulder structure shows at ca. 0.1 Å (see Figure 2). The whole profile of the Coulomb hole is very sensitive to the basis set. Increasing the number of basis functions improves the description of interelectronic cusps, shifting the hole to shorter electron-electron distances. For small basis sets, including only S and P functions, the shoulder is actually a minimum, in accord with the results presented by Cioslowski and Liu that also employed only S and P functions.[11] The shoulder structure, as reported by Bunge and Cioslowski[11,37] shows using S, P, and D functions. Augmenting with F functions does not produce a large change, and the addition of G functions barely changes the Coulomb hole, thus suggesting that the presence of the shoulder is not due to a basis set completeness problem (see Figure 2). Actually, the presence of the shoulder structure was also reported using Monte Carlo calculations.[53] The shoulder always appears when we use basis sets with enough flexibility to afford a correct description of core electrons, indicating that those electrons are causing the shoulder. The role of core orbitals is confirmed by the corresponding frozen-core CISD (fc-CISD) calculations which do not show any shoulder structure (see Figure S5 in the Supporting Information). Correlation effects on the $K$-shell of Ne have been previously studied by Buijse and Baerends. They used a modified version of the Coulomb hole proposed by Ros[50] based on conditional probability densities. They showed that core-electron correlations have a major impact on the Coulomb hole plots for Ne when the reference electron is located on the $K$-shell.[2]

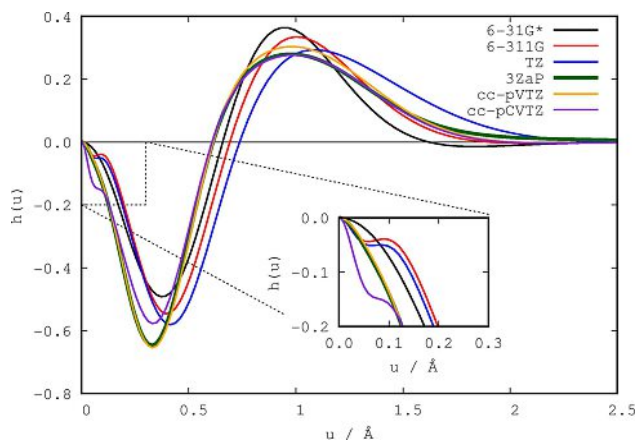

**Figure 1.** The CISD Coulomb hole of Ne for some selected basis sets.

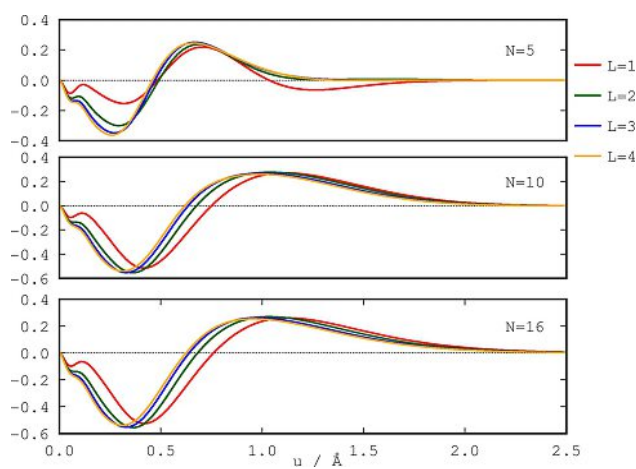

**Figure 2.** The CISD Coulomb hole of Ne for some even-tempered basis sets.

[a]  Dr. M. Rodríguez-Mayorga, Dr. E. Ramos-Cordoba, Prof. X. Lopez, Prof. J. M. Ugalde, Dr. E. Matito
Kimika Fakultatea, Euskal Herriko Unibertsitatea (UPV/EHU), and Donostia International Physics Center (DIPC), P.K. 1072, 20080 Donostia, Euskadi, Spain
E-mail: eloy.raco_at_gmail.com
E-mail: ematito@gmail.com

[b]  Dr. M. Rodríguez-Mayorga, Prof. M. Solà
Institut de Química Computacional i Catàlisi (IQCC) and Departament de Química, University of Girona, C/ Maria Aurèlia Capmany, 69, 17003 Girona, Catalonia, Spain

[c]  Dr. E. Matito
IKERBASQUE, Basque Foundation for Science, 48011 Bilbao, Euskadi, Spain

### 3.3. The Origin of the Shoulder

In this section we analyze the reasons for the existence of the shoulder in the Coulomb hole of Ne. We have already established that the correlation of core electrons is responsible for it. Let us now consider the importance of different configurations by removing some of them from the CISD expansion calculated with the 9SP basis set. In Figure 3 we have plotted the Coulomb hole generated with this wave function and other wave functions in which we have truncated the CISD expansions including only some excitations from the $1s$ orbital (see caption of Figure 3 for more details). The truncation of the determinant expansion has been performed after a standard CISD calculation, normalizing the resulting wave function after the truncation. The CISD expansion in which we have removed all the excitations from the core orbital except the single excitations (CISD(nc)+A in Figure 3) produces a Coulomb hole that is virtually identical to the fc-CISD one. The double excitations involving only one electron in $1s^2$ produce likewise a Coulomb hole qualitatively similar to the fc-CISD wave function (CISD(nc)+B). Among the double excitations the most important ones are those exciting simultaneously both $1s^2$ electrons as evidenced from the shoulder structure of the Coulomb hole of the CISD wave function where only these excitations from the core orbital are retained (CISD(nc)+C). A detailed analysis of the double excitations from the $1s$ orbital shows that the preferred virtual orbitals are $4s$, $5s$, $5p$, and $6p$ (see CISD(nc)+D Coulomb hole in Figure 3). These results have been qualitatively confirmed with the CISD/16SPDFG wave function (see Figure S9 in the Supporting Information).

Figure 4 plots the Coulomb hole for CISD expansions that only include the HF configuration and some chosen configurations involving excitations from the $1s$ orbital. Unlike the previous CISD expansions, these ones only include correlation effects due to the core electrons in Ne and, therefore, should reflect the importance of certain configurations in retrieving the shoulder. The inclusion of double excitations from the core orbitals gives rise to a hole structure (see HF+E in Figure 4) that is responsible for the shoulder structure of the complete CISD expansion. From this plot is also evident that double excitations and particularly those involving $4s$, $5s$, $5p$, and $6p$ are mostly responsible for the shoulder structure.

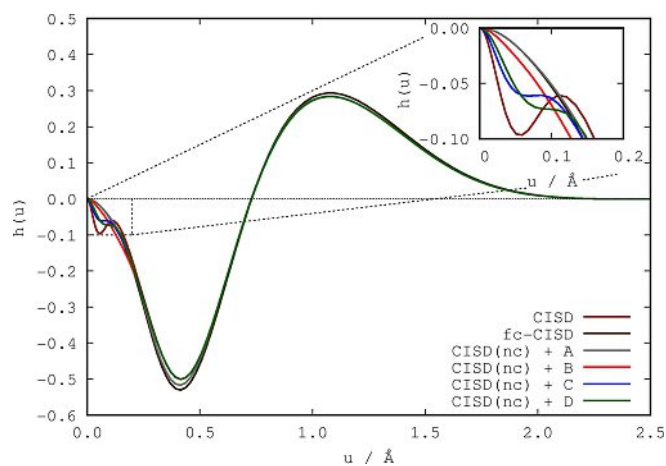

**Figure 3.** The CISD/9SP Coulomb hole in terms of several expansions. fc-CISD calculations were obtained from a CISD calculation in which no excitations from core orbitals were allowed, whereas CISD(nc) is a regular CISD calculation in which the configurations involving excitations from the core orbital have been removed *a posteriori*. A–C are groups of configurations including various excitations from the core orbital: (A) single excitations, (B) double excitations involving only one electron in the core orbital, (C) double excitations involving the two electrons in the core orbital excited to one single orbital, and (D) double excitations involving the two electrons in the core orbital excited to orbitals $4s$, $5s$, $5p$, and $6p$. After removal and addition of these configurations, the expansion coefficients have been rescaled to attain the normalization of the wave function.

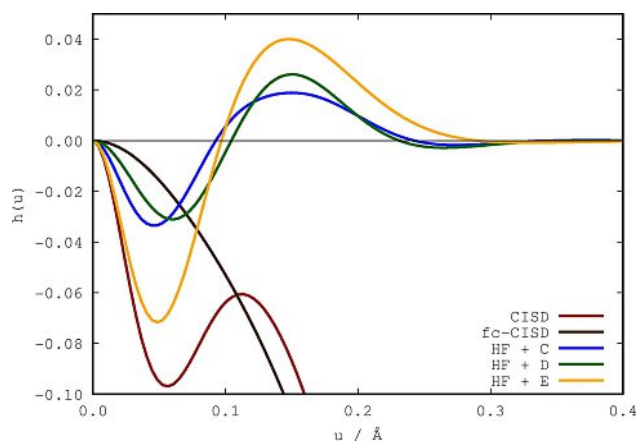

**Figure 4.** The CISD/9SP Coulomb hole in terms of several expansions. The groups of configurations included involve excitations from the core orbitals to some particular virtual orbitals (see the caption of Figure 3 for C and D). The E group includes configurations involving double excitations from $1s$ to all virtual orbitals.

Thus far, we have firmly established the presence of the shoulder in the Coulomb hole of Ne, which is due to the

   

electron correlation of the core electrons. In the following, we will analyze how the correlation affects the electronic structure of Ne and the particular role that the core electrons play in this context using the CISD/16SP wave function. First of all, we will consider the shell-structure of Ne. There has been some controversy in the literature concerning the descriptor that should be employed to identify the shell structure and shell numbers in atoms,[23,32,54] in our opinion, the one-electron potential (OEP) of Kohout being the most robust suggestion made thus far.[22] According to the OEP, we find that the radius of the $K$ shell does not change upon inclusion of electron correlation effects ($r_K = 0.138$ Å) and the $K$-shell number only increases $3 \cdot 10^{-3}$ electrons due to correlation ($n_K^{HF} = 2.0019$ e.). Therefore, according to the shell structure determined by the OEP, we conclude that electron correlation does not cause an expansion or contraction of the $K$ shell, but a small reorganization within the $K$ shell.

We have also checked the convergence of the electron-electron repulsion and the electron-nucleus attraction to see how these quantities are affected by the frozen-core approximation. These energy components show a convergence pattern that alternates fc-CISD results with CISD results, suggesting that wave functions that do not show the shoulder structure do not converge these properties differently (see Figures S6 and S7). Conversely, as one could expect, we have found that the intracule of the pair density at the coalescence point divided by the charge-concentration index, $\int \rho^2(\mathbf{r}) d\mathbf{r}$,[55,60] is affected by the inclusion of core correlation (see Figure S8).

Finally, let us assess the type of correlation affecting the shoulder structure. We will use our recently introduced separation of dynamic and nondynamic correlation scheme[42,44] that we have lately extended to separate the correlation in Coulomb holes.[62] In Figure 5 we can see that the short-range part of the Coulomb hole corresponds mostly to dynamic correlation and that the shoulder structure is also present in this part of the Coulomb hole.

### 3.4. Second-row Atoms and Molecules

In this section we investigate whether the shoulder is a feature of the Coulomb hole of Ne or other second-row atoms also show a signature of core-electron correlation at short interelectronic distances of their holes. From previous studies,[34,36] it is known that the Coulomb holes of He and Li do not present such a shoulder. For the rest of second-row atoms in their ground states, a shoulder or a minimum is always obtained as we show in Figure 6 (the complete holes can be found in

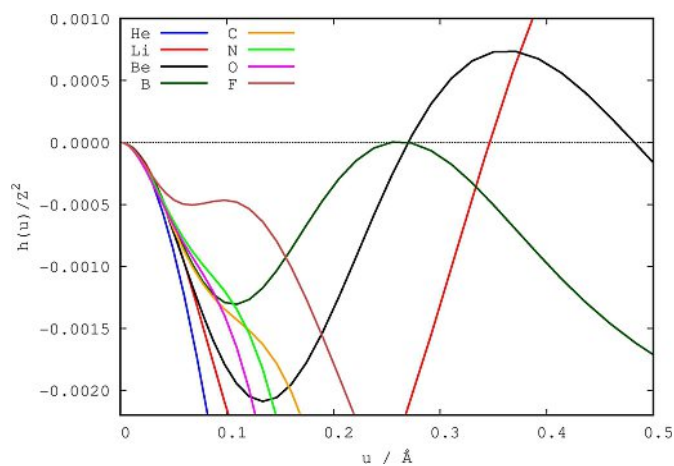

**Figure 6.** Zoom of the short-range CISD/6-311G* Coulomb holes of the second-row atoms.

Figure S10). In Figure 6 we plot the Coulomb hole divided by the square of the atomic charge ($Z^2$) in order to make all the holes fit in the same scale. The Coulomb holes reported for the open-shell systems were obtained using an unrestricted formalism (i.e. they correspond to the difference between the UCISD and the UHF radial intracule densities). Our study reveals that for Be, B, and F atoms, a minimum of the Coulomb hole is observed, while for C, N, and O atoms, a shoulder is produced. The shoulder or minimum vanish when they are calculated employing a fc-CISD wave function (see Figure S11), proving that the features observed correspond to correlation effects of the core electrons. The analysis of the OEP reveals that in all cases the radius of the $K$ shell does not change upon inclusion of electron correlation effects, and only an internal small reorganization within the $K$ shell is produced (see Table S4 for more details).

## 4. Conclusions

We have analyzed the Coulomb hole of Ne from highly-accurate CISD wave functions. Our energy estimates have been obtained from a two-fold extrapolation of optimized even-tempered basis sets and compare well with the best estimates available in the literature (we recover 97% of the correlation energy of Ne). We have confirmed the existence of a shoulder in the short-range region of the Coulomb hole of the Ne atom, which is due to the correlation of the core electrons in the *K* shell. Double excitations from the core orbital give rise to the most important configurations in the CISD expansion that contribute to the shoulder. The shoulder is due to an internal reorganization of the *K* shell, where electrons are pushed towards the *K*-shell boundary. The correlation nature of the shoulder is dynamic, as one would expect. This feature is very sensitive to the basis set in the core region. Finally, we have proven that for the rest of second-row atoms, except Li, a shoulder or a maximum in the short-range region of the Coulomb hole is obtained, which is due to the correlation of the core electrons in the *K* shell. In all cases, the shoulder or the minimum corresponds to an internal reorganization of the *K* shell.

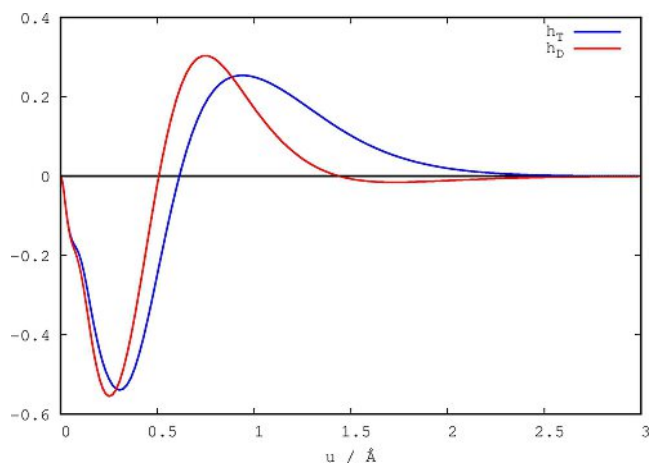

**Figure 5.** The dynamic part ($h_D$) and the total ($h_T$) Coulomb hole of Ne at the CISD/16SPDFG level of theory.

## Acknowledgements

This research has been funded by the Spanish MINECO/FEDER Projects CTQ2014-52525-P (E.M.), CTQ2017-85341-P (M.S.), CTQ2015-67608-P (X.L.), CTQ2015-67660-P (J.M.U.) and EUIN2017-88605 (E.M.), the Basque Country Consolidated Group Project No. IT588-13, and the Generalitat de Catalunya (Project 2017SGR39, Xarxa de Referència en Química Teòrica i Computacional, and the ICREA Academia 2014 prize (M.S.)). The FEDER grant UNGI10-4E-801 (European fund for Regional Development) has also supported this research. M.R.M. acknowledges the Spanish Ministry of Education, Culture and Sports for the doctoral grants FPU-2013/00176. E.R.C. and E.M. acknowledge funding from the European Union's Horizon 2020 research and innovation programme under the Marie Sklodowska-Curie grant agreement (No. 660943). The authors acknowledge the computational resources and technical and human support provided by the DIPC (especially to Daniel Franco and Diego Lasa) and the SGI/IZO-SGIker UPV/EHU.

## Conflict of Interest

The authors declare no conflict of interest.

**Keywords:** Coulomb hole · Neon · electronic structure · benchmark · electron correlation · core electrons